\documentstyle[aps,prb,epsfig]{revtex}
\begin{document}
\date{\today}
\title {Temperature Dependence and Mechanisms of Vortex Pinning by Periodic Arrays of Ni Dots in Nb Films}

\author{J. I. Mart\'\i n \cite{oviedo} and M. V\'elez \cite{oviedo}}
\address{Depto. F\'\i sica de Materiales, F. F\'\i sicas, Universidad
Complutense, 28040 Madrid, SPAIN}
\author{A. Hoffmann}
\address{Physics Department, University of California-San Diego, La Jolla, CA
92093, USA \\ and Manuel Lujan Jr., Neutron Scattering Center, Los Alamos National Laboratory, Los Alamos, NM 87545, USA}
\author {Ivan K. Schuller} 
\address{Physics Department, University of California-San Diego, La Jolla, CA
92093, USA}
\author{J. L. Vicent}
\address{Depto. F\'\i sica de Materiales, F. F\'\i sicas, Universidad
Complutense, 28040 Madrid, SPAIN}
                                       
\maketitle

\begin{abstract} Pinning interactions between superconducting vortices in Nb and magnetic Ni dots were studied as a function of current and temperature to clarify the nature of pinning mechanisms. A strong current dependence is found for a {\it square} array of dots, with a temperature-dependent optimum-current for the observation of periodic pinning, that decreases with temperature as $(1-T/T_C)^{3/2}$. This same temperature dependence is found for the critical current at the first matching field with a {\it rectangular} array of dots. The analysis of these results allows to narrow the possible pinning mechanisms to a combination of two: the interaction between the vortex and the magnetic moment of the dot and the proximity effect. Moreover, for the {\it rectangular} dot array, the temperature dependence of the crossover between the low field regime with a {\it rectangular} vortex lattice to the high field regime with a {\it square} configuration has been studied. It is found that the crossover field increases with decreasing temperature. This dependence indicates a change in the balance between elastic and pinning energies, associated with dynamical effects of the vortex lattice in the high field range.   
\end{abstract}
\pacs{}

\narrowtext
\section{INTRODUCTION}
Vortex dynamics is an active research field due to the richness and variety of the physics it presents. The superconducting material, defect structures, temperature and field are the main factors that strongly affect the behavior of the vortices and give rise to complex phase diagrams \cite{1,2}. Moreover this field has attracted much interest due to the hope that understanding vortex motion and pinning interactions may help improve current carrying characteristics of superconducting materials for eventual applications.

The recent development of techniques such as e-beam \cite{3,4}, x-ray \cite{5} or laser-interference lithography \cite{6} has opened up the possibility to study the interaction of the vortex lattice with periodic arrays of submicrometric pinning centers. Thickness modulations \cite{7}, arrays of holes \cite{8,9} or arrays of metallic and insulating dots \cite{10,11,12,13,14,15} give rise to interesting commensurability effects and stabilization of new geometrical configurations of the vortex lattice \cite{16}.

In particular, synchronized pinning by metallic and magnetic dots in superconducting thin films has been studied as a function of array geometry \cite{11,12,13,16} (triangular, kagome, square or rectangular), dot size \cite{15}, dot material \cite{12,15}, dot magnetic state \cite{14}, etc. This periodic pinning depends strongly on the vortex lattice properties (i.e. intervortex interactions, elastic energy and vortex velocity) and the dots characteristics (i.e. size, interdot distance, magnetic properties, etc). An additional complication is caused by the presence of random defects in the superconducting film that compete with the synchronized effect \cite{11}. Because of these, a clear description of the main pinning mechanisms by the magnetic dots is still missing.

To address this important issue, we have studied the synchronized pinning in Nb films with square and rectangular arrays of Ni dots. Different regimes are observed in the field dependence of the resistivity and critical current. In addition, with rectangular arrays of pinning centers, the temperature dependence of the vortex lattice reconfiguration from rectangular to a square geometry is studied. This helps discriminate among the different pinning mechanisms proposed \cite{12,14,15} (proximity effect, stray field, dot permeability, corrugation effects,... ).

The paper is organized as follows: in section II the sample preparation is described, in section III the current and temperature dependence of periodic pinning is presented, in section IV the mechanisms and the temperature dependence of periodic pinning in samples with rectangular arrays is discussed and the conclusions are presented in section V. 

\section{EXPERIMENTAL}
Ordered arrays of submicrometric Ni dots are fabricated by e-beam lithography combined with a lift-off technique as described elsewhere \cite{17}. Briefly, a Si substrate is covered by a PMMA resist layer where the desired pattern is written in a Scanning Electron Microscope (SEM). Then the sample is developed and a Ni layer is grown on the resist template. This is followed by a lift-off process in acetone to obtain the magnetic dot array. Figure 1 shows an SEM micrograph of one of these arrays, with typical dot dimensions of 40 nm thickness and 250 nm diameter. The dots are arranged either in square or rectangular arrays as in Fig. 1 with lattice parameters in the 400 nm - 900 nm range and extend over 50 $\mu$m $\times$ 50 $\mu$m areas. On top of the magnetic dot array, a 100 nm thick Nb film is deposited by sputtering. For the transport measurements a 40 $\mu$m bridge is defined by standard optical lithography and reactive ion etching. The samples present a metallic normal state behavior with a superconducting transition temperature in the $T_C$  = 8.0 - 8.5 K range. 

Transport measurements are performed in a helium cryostat, with the applied field perpendicular to the substrate plane. In the samples with rectangular (square) arrays of dots the current is applied parallel to the long side (one side) of the array cell.

\section{RESULTS}
A classical tool in the study of pinning interactions is to compare the field and temperature dependence of the pinning force with theoretical models \cite{1,18,19}. However, in the case of periodic pinning the situation is complicated by the competition between random and ordered pinning centers \cite{11}, and the existence of many possible vortex lattice configurations. At the different matching fields single, interstitial and/or multiquanta vortices may be present \cite{9,20,21}.

This complex behavior is illustrated in Fig. 2, which shows the voltage {\it vs.} field for a Nb film with a (400 nm $\times$ 400 nm) square array of Ni dots. Each curve is measured with a different value of the driving transport current. In a limited current range, $2.5 \times 10^{3}$ A/cm$^{2} <  J < 7.5 \times 10^{3}$ A/cm$^{2}$, the $V(B)$ curves show clear resistivity dips associated with synchronized pinning. These minima appear at constant $\Delta$$B$ = 130 G field intervals, which corresponds to a vortex density $n_{v} = B/\Phi_{0} = 6.3 \times 10^{8}$ cm$^{-2}$, where $\Phi_{0}$ is the quantum of flux. This is in very good agreement with the pinning centers density $n_{P} = 1/(400$ nm $\times$ 400 nm) = $6.25 \times 10^{8}$ cm$^{-2}$ of the square dot array. However, for low and high current densities, there is a monotonic dependence of the voltage increasing with magnetic field. Moreover, for this sample, no indication of periodic pinning is found in the field dependence of the critical current density.

This kind of driving current dependence is a clear indication of a change in the dominant pinning centers from the random defects in the Nb film to the ordered dot array. This is related to dynamical ordering in the vortex lattice at high vortex velocities \cite{11,22}. Therefore, the field and temperature dependence of the critical current is affected by random pinning and complicates the interpretation of the pinning mechanism by the magnetic dots. However, it is interesting to analyze instead the behavior of the optimum current $J_{opt}$ for the observation of periodic pinning, defined as the current where the highest number of minima in the resistivity vs field curves could be observed. Figure 3 shows that $J_{opt} \propto (1-T/T_C)^{3/2}$. However, the available current range $\Delta$$J_{opt}$ for the observation of periodic pinning becomes narrower the lower the temperature. For example $\Delta$$J_{opt}/J_{opt}$ = 1 at $T = 0.99T_C$, while $\Delta$$J_{opt}/J_{opt}$ = 0.16 at $T = 0.97T_C$. This suggests that, as the temperature is reduced, pinning by the periodic array becomes weaker than pinning by random defects.

Periodic pinning is also observed for a Nb film with a rectangular (400 nm $\times$ 625 nm) array of Ni dots, as shown in fig. 4, where the critical current ($J_C$) and the resistivity ($\rho$) are plotted as a function of magnetic field at $T = 0.98T_C$. In this case, two field regimes are found with two different field spacings between the periodic minima \cite{16}. At low fields, the minima appear at field intervals $\Delta$$B_{low}$ = 81 $\pm$ 2 G that corresponds to a vortex density $n_{v} =$ (3.9 $\pm$ 0.1) $\times$ 10$^{8}$ cm$^{-2}$ which is the same as the density of pinning centers $n_{P} = 1/(400$ nm $\times$ 625 nm) = $4 \times 10^{8}$ cm$^{-2}$. In the high field regime the distance between minima is $\Delta$$B_{high}=$ 112 $\pm$ 5 G that corresponds to a vortex lattice with lattice parameter $a_{0} = (\Phi_{0}/B)^{1/2} =$ 430 $\pm$ 10 nm, very similar to the short side (400 nm) of the rectangular array cell. This suggests that, in the low field regime, the vortex lattice is distorted into a rectangular configuration, while in the high field region the vortices form a square lattice that is pinned when it matches the short side of the rectangular array cell. The behavior in the two field regimes is different, not only in the shape and position of the periodic pinning minima, but also in the background dissipation at fields outside from the matching condition, i.e. there is an overall increase in the critical current (decrease in the resistivity) in the high field region. This can be related to the vortex lattice becoming stiffer at higher fields. As vortices interactions become stronger, the lattice cannot easily flow around pinned vortices so that the whole lattice is affected by the presence of the Ni dots at all fields even outside the matching condition.

The critical current has also a different temperature dependence in these two field regimes (see Fig. 5, where $J_C(T)$ is plotted for the fields corresponding to the first low and high field minima of the resistivity vs field curves). The temperature dependence of the critical current at the first low field minimum can be well fitted by a $(1-T/T_C)^{3/2}$ dependence, as shown in the inset of Fig. 5. This is the same power law found for $J_{opt}$ in the sample with the square lattice of dots, indicating that we are observing the same vortex-pinning center interaction in both cases.

On the other hand $J_C(T)$ has a different behavior for the first high field minimum. It is almost the same as the low field minimum $J_C(T)$ for $T >$ 8.1 K,  and then it falls below as the temperature is reduced. This indicates that  the vortex lattice-array of dots interaction is changing at this field position as a function of temperature. This change in the high field region as a function of temperature is also found in the magnetoresistance curves shown in Fig. 6, where the crossover between the high and low field regimes is found to have a weak temperature dependence. Close to $T_C$, the crossover occurs between the $3^{rd}$ and $4^{th}$ minima, whereas at lower temperatures it happens at a field position between the $4^{th}$ and $5^{th}$ minima. Thus, these results indicate a change as a function of temperature in the balance between elastic and pinning energies that govern this crossover as discussed below.

\section{DISCUSSION}
\subsection{Pinning Mechanisms}
There are several kinds of interactions between vortices and the Ni dots that have been proposed to account for the observed pinning effect. Some of them take into account the ferromagnetic character of the Ni dots, such as the reduction in the order parameter close to the dot by a magnetic proximity effect \cite{12,15} or by the dot stray field \cite{10,14}, the higher permeability at the dot \cite{12,15}, or the interaction between the vortex magnetic field and the magnetic dipole of the dot \cite{23}. There are also other possibilities that could contribute to the periodic pinning effect not of magnetic origin such as the periodic corrugation of the Nb film grown on top of the Ni dot array \cite{12}.

A good summary of the field and temperature dependence of the pinning force for a variety of non-ferromagnetic pinning centers can be found in the classical paper by Dew-Hughes \cite{19}. However none of those mechanisms can account for the behavior of the Ni dots (i.e. provide the observed $J_C(T)$  dependence with right pinning center characteristics). For example, pinning by point defects (i.e. with all three dimensions smaller than the penetration depth $\lambda$ and intervortex spacing as is the case here) gives rise to a temperature dependence in the pinning force linear in $(1-T/T_C)$ inconsistent with our observations. Therefore it is worth analyzing in some detail the pinning mechanisms based on the ferromagnetic character of the dots.

First, we consider the change in vortex magnetic energy due to the higher permeability of a dot with radius $r_{dot}$ and thickness $t_{dot}$. In this case, the pinning energy ${\varepsilon}_{P}$ is given by
\begin{equation}
{\varepsilon}_{P} = {{{\mu}_{0}} \over 2} {{\int}_{Vdot}}  \chi H^2 dV = {{{\Phi_0}^{2}t_{dot} \chi} \over {4 \pi \mu_{0} \lambda^4}} {\int_{0}}^{rdot} {[K_0({r \over \lambda})]}^2 rdr ,
\end{equation}
where $\chi$ is the Ni dot susceptibility, $\mu_0$ is the free space permeability, and the magnetic field of a vortex is \cite{24}
\begin{equation}
H(r) = {{\Phi_0} \over {2 \pi \lambda^2 \mu_0}} K_{0}({r \over \lambda}) ,
\end{equation}
with $K_0(x)$ the zeroth-order Hankel function. $\varepsilon_P$ is the gain in energy when the vortex is moved from infinite to the dot. Taking into account that the vortex field is spread over a distance $\lambda$, this change in energy will also occur in an interval of the order of $\lambda$. Therefore, in a first approximation, the pinning force can be estimated as
	\begin{equation}
F_P \propto {{\varepsilon_P} \over {\lambda}} = {{{\Phi_0}^2 t_{dot} \chi} \over {4 \pi \mu_0 \lambda^5}} {\int_0}^{rdot} {[K_0({r \over \lambda})]}^2 rdr .
\end{equation}
In this temperature range $r_{dot}$ is larger than the coherence length $\xi$ and smaller than the penetration depth $\lambda$, i.e. $\xi < r_{dot} < \lambda$  (typical values for Nb films at $T = 0.98T_C$ are $\xi$ = 90 nm and $\lambda$ = 320 nm while $r_{dot}$ = 130 nm). Then, the integral of $K_0(x)$ function in eq. (3) is weakly temperature dependent (with only a 10\% change in the temperature range of the experimental data in Fig. 5) and, therefore, the temperature dependence of the pinning force is dominated by the factor 1/$\lambda^5$ so that
	\begin{equation}
F_P \propto (1-T/T_C)^{5/2}.
\end{equation}

These results can be compared with $J_C(T)$ at the first minimum for the rectangular array (shown in Fig.5). At this field position the vortex lattice exactly matches the array of dots with probably one vortex pinned at every Ni dot. Therefore, $J_C$ is directly associated with the pinning force $F_{P} = J_{C}\times \Phi_{0}$ between one vortex with one magnetic dot. Then, the observed experimental behavior ($J_C \propto (1-T/T_C)^{3/2}$) is inconsistent with eq. (4). Thus the higher permeability of the dot does not provide a viable mechanism.

Another possible origin of pinning by ferromagnetic particles is the interaction between the vortex field and the Ni dot magnetic dipole \mbox{\boldmath${\mu}$} \cite{23}. The interaction energy between the vortex and a magnetic particle of moment \mbox{\boldmath${\mu}$} given by
\begin{equation} 
\varepsilon_P = {\bbox{\mu}} \cdot {\bf H} = {{\Phi_0 \mu_z} \over {2 \pi \lambda^2 \mu_0}} K_0({r \over \lambda}) .
\end{equation}
This energy gives rise to an attractive force
\begin{equation} 
F = - \mid \nabla ({\bbox{\mu}} \cdot {\bf H}) \mid = -{{\Phi_0 \mu_z} \over {2 \pi \lambda^3 \mu_0}} K_1({r \over \lambda}) ,
\end{equation}
where $\mu_z$ is the component of the magnetic moment parallel to the vortex field and $K_1(x)$ is the first-order Hankel function. As $r \rightarrow 0$, $K_1({r/\lambda})$ increases as $({r/\lambda})^{-1}$ with a cutoff at $r = \xi$ due to the vortex core. This corresponds to a maximum value of the pinning force
	\begin{equation}
F_P = {{\Phi_0 \mu_z}\over {2 \pi \mu_0 \xi \lambda^2}} .
\end{equation}
Close to $T_C$, using the Ginzburg-Landau expressions for $\lambda$ and $\xi$, the temperature dependence of this pinning force is given by
	\begin{equation}
F_P \propto {1 \over {\xi \lambda^2}} \propto (1-T/T_C)^{3/2} ,				\end{equation}
which is compatible with the experimental results of Fig. 5. For a Ni dot with dimensions $t_{dot}$ = 40 nm and $r_{dot}$ = 130 nm $\mu_{z}$ can be estimated as $\mu_z = \mu_0 \chi_z H_z \pi r_{dot}^{2} t_{dot} = 3 \times 10^{-23}$ Tm$^3$. Here $\chi_z$ is calculated using the formulas for an oblate ellipsoid \cite{25} as $\chi_z = 1/(N_a - N_c)$ = 1.6, with $N_a$ and $N_c$ the demagnetizing factors, and a typical value of  $H_z$ = 100 Oe has been considered. This value of $\mu_z$ gives $F_P = 10^{-7}$ dyn at 0.98$T_C$.

Finally, another possible pinning mechanism is provided by the presence of normal regions around the dot caused by either the dot stray field or the ferromagnetic proximity effect. Stray fields cannot be very relevant since the observed pinning effect is independent of magnetic history and shows no hysteresis effects. On the other hand, a clear indication of proximity effect is found in the strong $T_C$ depression of a Nb film grown on a Ni layer prepared under identical conditions as the Nb films on top of the Ni dots arrays used in this work \cite{26}. Morevoer, the presence of normal regions around the dot is consistent with the superconducting wire network behavior found in samples with very small dot separation \cite{15}.

The pinning energy of a normal inclusion with radius larger than $\xi$ inside a superconducting film of thickness $t_{film}$ can be estimated \cite{1} as
\begin{equation}
\varepsilon_P = {1 \over 2} \mu_{0}{H_{C}}^2 \pi \xi^{2}t_{film} = {{{\Phi_0}^ 2 t_{film}}\over{16\pi {\kappa}^2 \mu_0 {\xi}^2}} ,				\end{equation}
where $H_C$ is the thermodynamic critical field and $\kappa$ the Ginzburg-Landau parameter. The pinning force is given by
\begin{equation}
F_P = {{\varepsilon_P}\over{\xi}} \propto {1\over{\xi^3}} \propto (1-T/T_C)^{3/2},		
\end{equation}
which is also in good agreement with the observed $J_C(T)$ dependence in our samples. A more refined numerical calculation \cite{27} solving the two dimensional Ginzburg-Landau equations for this problem gives the same temperature dependence of the pinning force. This pinning mechanism provides a pinning force $F_P = 7 \times 10^{-8}$ dyn at 0.98$T_C$ for our samples, similar to the previous case. In both cases, this value is in good agreement with the order of magnitude of the pinning force by a magnetic dot $F_P = 10^{-7}$ dyn obtained earlier \cite{16} from the analysis of the vortex lattice reconfiguration in similar samples.

The temperature dependence of the critical current density is consistent with two pinning mechanisms that provide similar pinning force strength, both related with the ferromagnetic character of the dots. These two mechanisms, constitute opposite limits of the same problem. The first corresponds to a "dirty" interface between the Ni dot and the Nb film in which the dot does not perturb the superconducting properties of the Nb film, so that pinning occurs via a magnetic interaction with an unperturbed vortex. The second is that of a perfectly "clean" interface in which the proximity effect with the Ni dots creates a normal region in the Nb film and pinning occurs only by reduction in the vortex core energy. Therefore, in a real sample, the pinning interaction between a vortex and a Ni dot must be a combination of both mechanisms. These results are in good agreement with previous works that show strong enhancement of the periodic pinning in samples with magnetic dots compared to samples with non-magnetic dots \cite{12,14,15}.

\subsection{Temperature dependence of  pinning by rectangular arrays}
The elastic properties of the vortex lattice can be probed by the study of periodic pinning with non-isotropic arrays of defects. In particular, an overall deformation in the vortex lattice can be induced in order to match the geometry of a rectangular array of dots. This rectangular configuration of the vortex lattice is stable at low fields but at higher field there is a crossover to a more isotropic square geometry. The energy balance that governs this crossover is 
\begin{equation}
\Delta E = E_{rectangular} - E_{square} = \Delta E_{elastic}(rectangular-square) - \Delta E_{Pinning}, 	
\end{equation}
where $\Delta E_{elastic}(rectangular-square)$ is the increase in elastic energy due to the rectangular distortion relative to the square vortex lattice and $\Delta E_{Pinning}$ is the gain in pinning energy of the vortices at the magnetic dots. $\Delta  E_{elastic}(rectangular-square)$ becomes bigger as vortices interactions become stronger at higher magnetic fields, and induces the crossover between the low and high field regions \cite{16}.

The reduction in the crossover field for higher temperatures observed in Fig. 6, is an indication that the balance $\Delta E_{Pinning}/\Delta E_{elastic}$ decreases slightly with increasing temperature. To understand this change, we must consider the temperature dependence of the different terms. The increment in elastic energy considering vortices interactions in a bulk superconductor \cite{24} is
\begin{equation}
\Delta E_{elastic}(rectangular-square) = {{{\Phi_0}^2}\over{8\pi^2 \lambda^2}} [\sum_{rect} K_0({{r_{ij}}\over{\lambda}}) - [\sum_{square} K_0({{r_{ij}}\over{\lambda}})]  
\end{equation}
where $r_{ij}$ is the distance between vortices $i$ and $j$, and the sums are extended over two square or rectangular lattices of a density given by the magnetic field. In a thin film limit ($t_{film} \ll \lambda$), there is an additional interaction between vortices via the magnetic field outside the superconductor. Then, the $K_0(r_{ij}/\lambda)$ function in eq. (12) should be substituted by $H_0(2t_{film}r_{ij}/\lambda^2) - Y_0(2t_{film}r_{ij}/\lambda^2)$, with $H_0(x)$ the Struve function and $Y_0(x)$ the Bessel function of the second kind \cite{28}. The largest contributions to the sums of eq. (12) are given by the nearest and next nearest neighbors with $r_{ij} < \lambda$, where both $K_0(x)$ and $H_0(x)-Y_0(x)$ have the same dependence. Then, equation (12) can be written in both limits as
	\begin{equation}
\Delta E_{elastic}(rectangular-square) = {{{\Phi_0}^2}\over{8\pi^2 \lambda^2}}[\sum_{rect}[-log(r_{ij})] - \sum_{square}[-log(r_{ij})]].
\end{equation}
Close to $T_C$  the temperature dependence of this term can be approximated by
	\begin{equation}
\Delta E_{elastic}(rectangular-square) \propto {1\over{\lambda^2}} \propto (1-T/T_C)				
\end{equation}

The second term in eq. (11) is the difference between the pinning energy of the vortex lattice in the rectangular and the square configurations $\Delta E_{Pinning} = E_P$(rectangular) - $E_P$(square), where $E_P = \varepsilon_P f_P$, with $\varepsilon_P$ the pinning energy per dot and $f_P$ the fraction of vortices pinned at the dots for each particular field and vortex lattice configuration. In a first approximation, $E_P$(square) may be neglected because the fraction of pinned vortices in this configuration is much smaller than in the rectangular one. Assuming that at matching there is only one vortex per dot and neglecting the contribution of interstitial vortices, the gain in pinning energy of the vortex lattice is $\Delta E_{Pinning}= E_P$(rectangular) = $n_P \varepsilon_P$, with $n_P$ the density of pinning centers. Taking into account the relevant pinning mechanisms discussed above, $\varepsilon_P$ is given either by eq. (5) ($\varepsilon_P \propto 1/\lambda^2$) or by eq. (9) ($\varepsilon_P \propto1/\xi^2$) in both cases a linear dependence in $(1-T/T_C)$  close to $T_C$, i.e. 
\begin{equation}
\Delta E_{Pinning} = n_P \varepsilon_P \propto (1-T/T_C).
\end{equation}

Since the temperature dependence of eqs. (14) and (15) is the same, in a first approximation, the crossover field should be temperature independent. This is in agreement with the weak temperature dependence derived from the magnetoresistance data shown in fig. 6 and implies that a more refined calculation would be needed in order to account for the temperature change in the crossover. For example, for a proximity effect pinning mechanism the ratio $\Delta E_{Pinning}/\Delta E_{elastic} \propto \lambda^2/\xi^2$ so, by using the exact temperature dependence $\lambda(T) = \lambda(0)/\sqrt{1-(T/T_C)^{4}}$, results in a 6\% decrease in the crossover field from 8.0 K to 8.3 K. This is qualitatively in agreement with the experimental behavior although it is smaller than the change in crossover observed in Fig. 6 (a 15-20\% decrease in the same $T$ interval).

$E_P$(square) neglected above is the additional term that affects the temperature dependence of the gain in pinning energy, $\Delta E_{Pinning} = E_P$(rectangular)$ - E_P$(square).  In the high field region, the shape of the minima in magnetoresistance presents a strong temperature dependence, and they become much deeper as the temperature is increased (see fig. 6). This implies that the vortex lattice is more strongly pinned at the matching conditions, i.e. $E_P$(square) becomes more important close to $T_C$. In this high field regime, vortex-vortex interactions are strong enough to prevent distortion of the lattice by the rectangular array. This can give rise to a more complex behavior and factors such as the order in the vortex lattice, dynamical effects and the competition of periodic pinning with random defects become relevant. In general, the overall behavior in this high field region is quite similar to the sample with the square array discussed above. In that case, it was found that periodic pinning was favored for temperatures close to $T_C$, so that it could be observed in a wider current range. Therefore, the reduction in the crossover field must be related with this relative enhancement in pinning energy for the square vortex lattice at higher temperatures.

\section{CONCLUSIONS}
 In summary, we have analyzed the pinning mechanisms in Nb by ordered ({\it square} or {\it rectangular}) arrays of submicrometric Ni dots by studying the current and temperature dependences. The same $(1-T/T_C)^{3/2}$ temperature dependence is found in {\it square} arrays of dots for the optimum periodic pinning current $J_{opt}$ and in {\it rectangular} arrays for the critical current $J_C(T)$ at the first minimum in $\rho (H)$ (i.e. maximum in $J_C(H)$). This $T$ dependence is directly related to the pinning force between one vortex and one dot, since at this first minimum the vortex lattice exactly matches the array with one vortex pinned per every Ni dot. These results can be understood in terms of a combination of two pinning mechanisms related with the ferromagnetic character of the dots that provide the correct $T$ dependence and order of magnitude of the pinning force. 

For the {\it rectangular} arrays, two field regimes can be defined in the magnetoresistance data corresponding to two different geometrical configurations of the vortex lattice: {\it rectangular} at low fields and {\it square} at high fields. The crossover between both regimes is weakly temperature dependent and occurs at lower fields for higher temperatures. This indicates that the balance $\Delta E_{Pinning}/\Delta E_{elastic}$, that governs the {\it rectangular} to {\it square} crossover, decreases with increasing temperature. This is related to dynamical effects in the vortex lattice that induce an enhancement in the pinning energy of the {\it square} vortex lattice configuration at high temperature.

\acknowledgments
Work supported by Spanish CICYT (grant MAT99-0724), the US National Science Foundation, UC-CULAR and the Spanish-US Commission (grant 99002). We thank M. I. Montero, J. J\"{o}nsson, R. Sasik and J. P. Maneval for useful conversations.

\begin{figure}
\caption{SEM micrograph of a rectangular array of Ni dots fabricated by e-beam lithography.}
\end{figure}

\begin{figure}
\caption{Voltage {\it vs.} field for a Nb film with a square (400 nm $\times$ 400 nm) array of Ni dots at 8.35 K ($T_C$  = 8.43 K): A, $J = 10^4$ A/cm$^2$; B, $J = 7.5 \times 10^3$ A/cm$^2$; C, $J = 5\times 10^3$ A/cm$^2$; D, $J = 3.75 \times 10^3$ A/cm$^2$; E, $J = 2.5\times 10^3$ A/cm$^2$; F, $J = 1.25\times 10^3$ A/cm$^2$.}
\end{figure}

\begin{figure}
\caption{Temperature dependence of the optimal current for synchronized pinning for a Nb film with a (400 nm $\times$ 400 nm) array of Ni dots. Solid line is a linear fit to $J_{opt} \propto (1-T/T_C)^{3/2}$.}
\end{figure}

\begin{figure}
\caption{(a) Critical current and (b) resistivity as a function of magnetic field for a Nb film with a rectangular (400 nm $\times$ 625 nm) array of Ni dots at $T$ = 8.15 K.}
\end{figure}

\begin{figure}
\caption{Temperature dependence of the critical current for a Nb film with a rectangular (400 nm $\times$ 625 nm) array of Ni dots at the first low field minimum ($B$ = 81 G, empty symbols) and at the first high field minimum ($B$ = 324 G, filled symbols). Inset shows $J_C$  {\it vs}. $(1-T/T_C)^{3/2}$ at the low field minimum. Solid line is a linear fit to the data.}
\end{figure}

\begin{figure}
\caption{Field dependence of the resistivity for a Nb film with a rectangular (400 nm $\times$ 625 nm) array of Ni dots at (a) $T$ = 8.3 K and (b) $T$=8.0 K. (c) Position of the minima for: $T$ = 8.3 K, empty symbols; $T$ = 8.0 K, filled symbols. Solid lines are linear regressions.}
\end{figure}

\end{document}